%% file: paper.tex
\documentclass[runningheads]{llncs}
\usepackage{graphicx}
\usepackage{subcaption}
\usepackage{hyperref}
\usepackage{listings}
\usepackage{color}
\definecolor{dkgreen}{rgb}{0,0.6,0}
\definecolor{gray}{rgb}{0.5,0.5,0.5}
\definecolor{mauve}{rgb}{0.58,0,0.82}

\begin{document}
\title{The EuroSys 2020 Online Conference: Experience and lessons learned
}
%
%
\author{Angelos Bilas\inst{1,2} \and Dejan Kostic \inst{3} \and Kostas Magoutis\inst{1,2} \and Evangelos Markatos\inst{1,2} \and Dushyanth Narayanan \inst{4} \and Peter Pietzuch\inst{5} \and Margo Seltzer\inst{6}}
\authorrunning{}
%
\institute{
University of Crete, Greece
\and
FORTH-ICS, Greece
\and
KTH Royal Institute of Technology, Sweden
\and
Microsoft Research, UK
\and
Imperial College London, UK
\and
University of British Columbia, Canada
}
\maketitle              
\begin{abstract}
The 15th European Conference on Computer Systems (EuroSys’20)~\cite{10.1145/3342195} was organized as a virtual (online) conference on April 27-30, 2020. The main EuroSys’20 track took place April 28-30, 2020, preceded by five workshops (EdgeSys’20, EuroDW’20, EuroSec’20, PaPoC’20, SPMA’20) on April 27, 2020. The decision to hold a virtual (online) conference was taken in early April 2020, after consultations with the EuroSys community and internal discussions about potential options, eventually allowing about three weeks for the organization. This paper describes the choices we made to organize EuroSys'20 as a virtual (online) conference, the challenges we addressed, and the lessons learned.

\end{abstract}
\setlength{\parskip}{3pt}

\input{intro}
\input{implementation}
\input{analysis}
\input{challenges}

\section{Acknowledgments}
We would like to acknowledge the support of ACM, SIGOPS, and the
EuroSys Officers and Steering Committee throughout the organization,
to ACM and FORTH-ICS for providing Zoom licenses, to FORTH-ICS for
supporting the organization of the online conference in a number of
ways, and to the local arrangements team (Alison Manganas, Nikoletta
Palivakou), our Web chair and Zoom master (Antonis Krithinakis) and
volunteers (Jack Kolokasis, Christos Papachristos, Tasos Papagiannis,
Antonis Papaioannou, Manos Pavlidakis) for their support in the
successful implementation of virtual EuroSys’20.

%
%
%
\bibliographystyle{ieeetr}
\bibliography{bibliography}
\end{document}

%% file: intro.tex
\section{Introduction}
\label{ch:Introduction}

The 15th European Conference on Computer Systems
(EuroSys’20)~\cite{10.1145/3342195} was planned to be held in
Heraklion, Crete, Greece on April 27-30, 2020. 

\paragraph{Decision to go virtual:}
The unfolding of the COVID-19 global outbreak in early 2020 led
several private organizations and governments to consider limiting
international travel, and as a result it forced conferences including ASPLOS'20~\cite{asplos2020} and 
EDBT/ICDT'20~\cite{edbt-icdt2020} planned to
be organized in the March-April 2020 timeframe to consider alternative
plans.

In EuroSys'20, we spent the first week of March 2020 on intense
deliberations on how to respond to this urgent situation.  Given that
we were roughly two months away from the conference dates and there
were no firm projections about COVID-19, the decision was not clear.
Besides internal deliberations, we decided to also ask for the opinion
of the EuroSys community through an online survey on March 10-13.
Although several members of the community felt that a physical
conference cannot be replaced, the feedback we received on March 13
was pointing towards either canceling the physical event or holding a
smaller physical event.

We spent the following three weeks on internal brainstorming about
potential options, taking into account the potential impact on
organizational costs, logistics, and the general outlook of how the
COVID-19 situation was likely to evolve. All these were
  weighted against what constitutes the best service to the community:
  a virtual conference, a smaller event during the original dates with
  remote (online) options, or a (hopefully) typically-sized event at
  later dates. On April 5, we decided that holding EuroSys'20 as a
virtual (online) conference was the most reasonable course of action
and we announced this on the conference web page.

\paragraph{Preparation to go virtual:}
While we were aware that this was not going to be an easy task, since
none of us had organized a virtual conference before, we also felt
that this was an opportunity to explore new ideas. 

In EuroSys’20, we aimed to provide an interactive conference
experience, while also leveraging team collaboration and communication
tools for side-channel information exchange. In particular, EuroSys’20
combined:

\begin{itemize}
\item Synchronous Zoom\footnote{https://zoom.us} sessions, in webinar mode, i.e. limited rights
  for attendees, with streamed pre-recorded presentations
  and live Q\&A.
\item Asynchronous chat channels (Slack\footnote{https://slack.com}, Discord\footnote{https://discord.com}) for
  discussion among attendees, and also for coordination among
  organizers.
\item Synchronous virtual meeting rooms and hosted discussions during
  breaks, in the form of a hallway track via Discord.
\end{itemize}

These information channels would allow people to participate in
multiple ways, increasing their sense of participation.

\paragraph{During the conference:}
The main program broadcasted on Zoom sessions set the pace of the
conference program. Had asynchronous tools been used alone,
participants may have been unsure as to where “the action” happens at
any point in time.

Side channels, asynchronous or synchronous, via chat or
  voice, such as Slack and Discord, allowed information to flow at a
higher rate than synchronous sessions alone would allow.  A caveat
with using several communication channels is that the increased level
of information flow may potentially distract participants from the
live program (see Section~\ref{ch:Challenges}, lessons
learned); it can also overload organizers, as a large influx of
opinions, requests, feedback, etc. about the organization may distract
them from attending to higher priority events.

\paragraph{Aftermath:}
The feeling at the end of the conference and the feedback from the
attendees indicate that there was a sense of an online event with live
attendance and participation. Pre-recorded talks were of very high
quality and were received very well. The increased information flow
was not easy to manage. Interactive, voice side-channels are very
promising, although they were not used as much during the
event. Timezones are difficult to manage. Preparing a virtual
conference involves a lot of backstage work before and during the conference.

In the following sections, we describe in more detail the design and
implementation choices we made to organize EuroSys'20 as a virtual
(online) conference, an analysis of attendee feedback through a survey
that we run immediately following the conference, and finally the
lessons learned.

%% file: implementation.tex
\section{Design and implementation}
\label{ch:design}

In setting out to design and implement EuroSys’20, we benefited from
an ongoing discussion and report by the ACM Presidential Task Force on
‘What Conferences Can Do to Replace Face-to-Face
Meetings’~\cite{ptf-guide} and experience with previous online
conferences~\cite{asplos2020,edbt-icdt2020}. Preliminary experience
with the use of Zoom to deliver a synchronous online conference
program~\cite{edbt-icdt2020} showed that this can be done
effectively. In addition, it pointed out that it can also be enhanced
by putting operations staff in place to assist session chairs, by
serving rich content throughout the day, and by offering new ways for
synchronous and asynchronous interaction between participants.

\paragraph{Pre-recorded talks and live Q\&A:}
Zoom sessions were modelled after the typical sequence of events in a
physical conference, except that talks were shorter
(EDBT/ICDT~\cite{edbt-icdt2020} also featured shorter talks). We asked
authors to provide two versions of each talk: a short (3-5 minute)
version to be streamed at the conference, and a longer (10-12 minute)
talk to make available in advance so that attendees could watch in
preparation of the actual talk. While this put extra burden on
authors, it proved an important element of success as it allowed us to
experiment with talk duration and adapt in real time, which we
eventually did.

Each session included pre-recorded talks and live Q\&A. The first
session started with short videos (3mins) but after that we switched
to longer videos (12min) as the short videos did not seem to provide
adequate technical detail. The longer videos worked well until the end
of the conference. Most pre-recorded talk were of very high quality
and were received very well. The live Q\&A session was appreciated but
was not used as extensively as we would have hoped by attendees. The
feedback suggests that videos 11-14 minutes or longer are preferred.

Minor details, such as a recorded applause at the end of a talk, were
especially well received.

\paragraph{Tools and roles for each session:}
We decided to use Zoom webinars as a way to limit the rights of
participants and avoid potential misbehavior. We came up with the
following organizational roles for the production and broadcasting of
synchronous interactive content over Zoom:

\begin{itemize}
\item Content producers and graphics designers, tasked to create
  videos and other media to play before/after session, during breaks,
  sponsor slots, etc. This is in addition to the presentations that
  is streamed (live or pre-recorded during the conference);
\item Director, coordinates what content plays and when;
\item Zoom host (master), co-hosts (operators) to share screen and
  operate content flow, start/stop video sequence, Zoom participant
  registration, change of status (attendee $\leftrightarrow$ panelist
  $\leftrightarrow$ co-host);
\item A team of Zoom hosts were needed to serve as (a) stand-by hosts,
  in case of trouble with master host and (b) hosts for parallel
  workshops.
\end{itemize}

These roles would interface with the following traditional conference
roles:

\begin{itemize}
\item Organizers (general chairs, program chairs), participating in
  sessions as Zoom panelists;
\item Session chairs, participating in sessions as Zoom co-hosts (thus
  also Zoom panelists) so they could unmute participants during Q\&A;
\item Speakers, attendees that would be set to Zoom panelists for
  their session (for simplicity, we allowed them to be panelists for
  the entire day);
\item Student volunteers, assisting with Zoom sessions as Zoom hosts,
  making sure behind the scenes that progression within each session
  is smooth and reacting to unplanned changes and requests.
\end{itemize}

\paragraph{Training at large:}
It was obvious to us that we would need to perform significant
training, and embarked upon it early in the (time-constrained) preparation process. Our Zoom master led the
training of Zoom co-hosts so that they could serve as his backups, and
also as hosts of parallel workshop sessions. The Zoom master also
trained Session chairs to be able to operate as co-hosts during
sessions. Zoom co-hosts spun off and scheduled training tasks with
workshop organizers. The plan was also to invite speakers for dry
runs; however, this was not possible within the limited time we
had. In retrospect, it seems that it was not necessary either:
speakers seemed to find it easy to join as panelists and participate
in Zoom sessions for Q\&A, so such training may in fact be an optional
step in practice.

\paragraph{Instructions and guides:}
What helped significantly during training was the production of
textual user guides for session chairs and presenters:
\begin{itemize}
\item \url{https://www.eurosys2020.org/information-for-the-presenter/}
\item \url{https://www.eurosys2020.org/information-for-the-session-chair/}
\end{itemize}

\paragraph{Side channels:}
As we had mapped out what needed to be done with Zoom, we realized
that the community demanded also asynchronous teamwork platforms. In
particular, we received several requests for Slack early on based on
community experience with its use in the ASPLOS’20
conference~\cite{asplos2020} and in the CS Research and Practice
Slack~\cite{cs-slack}. To respond to this need we decided to implement
a simple Slack workspace for EuroSys’20. In the process, we understood
that we could use it also for internal coordination between the
several Zoom and conference roles; this coordination proved to be key
to the successful implementation of EuroSys’20.

A first implementation of the Slack workspace was straightforward, and
aided by experience with ASPLOS~\cite{asplos2020}. We set up regular
public communication channels for (a) Sessions (presentation videos,
Q\&A), (b) General (a lot of general activity, discussions, etc.), and
(c) Jobs.

During the conference, we realized that additional public channels
would increase people connection with the conference and improve their
experience:
\begin{itemize}
\item Sponsor channels (sponsor-related content);
\item Media (pictures, videos about the original location of the conference);
\item Stats (mainly participation figures, published soon after sessions ended); and
\item Posters.
\end{itemize}

Operating this workspace (producing content, responding to questions
and requests about organizational matters) required the involvement of
several staff members; to ensure responsiveness, we had to assign the
following tasks to a number of team members:
\begin{itemize}
\item Monitor and respond to organizational questions;
\item Regularly emit informational messages on \#general, such as sessions starting, etc.;
\item Produce statistics about the conference, such as attendance figures; and 
\item Produce media content (videos) to provide attendees with a visual experience.
\end{itemize}

\paragraph{Importance of private channels:}
To facilitate coordination and management between members of the
organization, we set up private chat channels for easy communication
and coordination between
\begin{itemize}
\item Session chairs and presenters for each session (\#mgmt-sessionX, \#mgmt-posters);
\item Organizers (general chairs, PC chairs) and session chairs (\#mgmt-session-chairs);
\item Local organizing team (general chairs \& Zoom/Slack/Discord hosts) (\#mgmt-local); 
\item Organizers and sponsors (\#mgmt-sponsors); and
\item Organizers and sponsor chairs (\#mgmt-sponsor-chairs).
\end{itemize}

These private channels proved important for the smooth operation of
the conference. The channels between session chairs and presenters
(where organizers also participated) proved to be invaluable in
maintaining contact early on and giving session chairs a channel to
communicate important information easily. We believe that this
coordination mechanism could be useful to some extent in the case of
physical conferences as well. The channel between organizers and
session chairs made it possible to adjust conference parameters in
real time and coordinate such changes seamlessly across the
organization. For instance, the organizers decided to change the
length of videos from short to long version at the end of session 1;
this change was agreed upon and communicated with everyone involved,
and applied without disruption in the program.

\paragraph{Facilitating sponsors:}
While not addressed in detail in this report, this model offered many
opportunities for sponsors to interact with attendees (on Zoom and
through Slack/Discord channels and Discord meeting rooms). We found
that it is important to engage sponsors early on, to communicate
opportunities, and to give them enough lead time to assign
representatives to participate in activities (Zoom and channels) and
to provide the necessary content.

%% file: analysis.tex
\section{Analysis of survey results}
\label{ch:Analysis}

To collect feedback from EuroSys’20 participants, we circulated a
questionnaire after the end of the conference, to which we received
100 responses. The composition of the questionnaire was influenced by
previous surveys of online conferences~\cite{asplos2020,edbt-icdt2020}. An analysis of the
collected responses appears below:

\subsection{Current occupation}
\begin{figure}[hbt!]
\centering
\includegraphics[width=0.7\columnwidth]{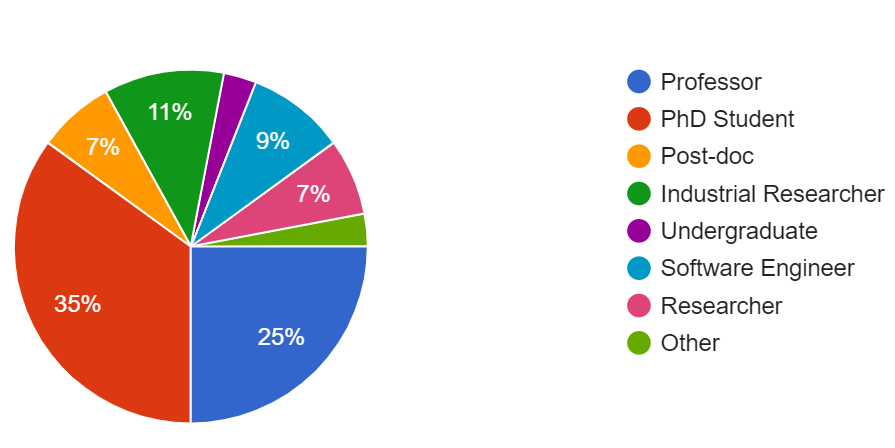}
\caption{\label{q1}Current occupation.}
\end{figure}

The largest percentage of respondents were PhD students (35\%), followed by
professors (25\%), industrial researchers (11\%), software engineers
(9\%), post-docs (7\%), and researchers (7\%) (Figure~\ref{q1}).

\subsection{Which continent were you on during the conference?}
\begin{figure}[hbt!]
\centering
\includegraphics[width=0.7\columnwidth]{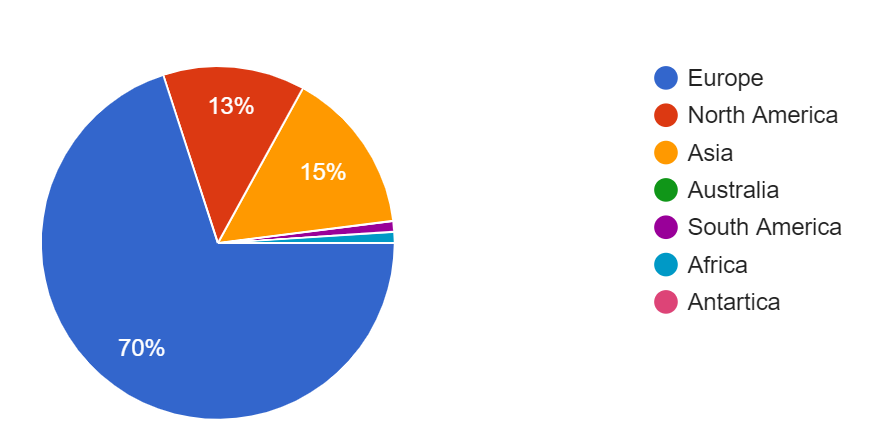}
\caption{\label{q2}Which continent were you on during the conference?}
\end{figure}

The majority of respondents were in Europe (70\%), followed by Asia
(15\%) and North America (13\%) (Figure~\ref{q2}).

\subsection{How many sessions did you attend?}
\begin{figure}[hbt!]
\centering
\includegraphics[width=0.6\columnwidth]{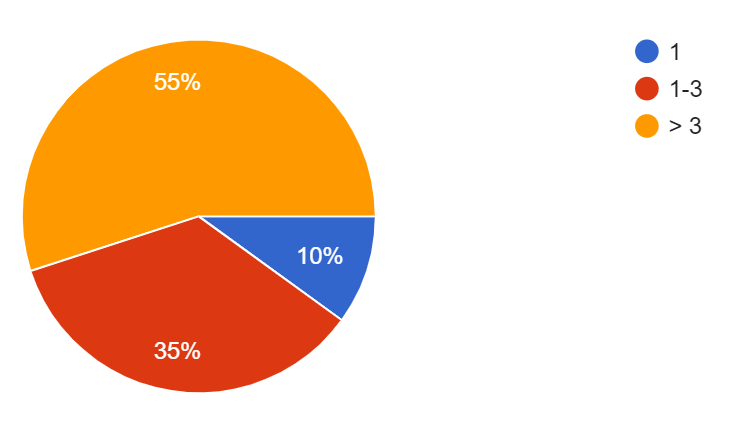}
\caption{\label{q3}How many sessions did you attend?}
\end{figure}

More than half of the respondents (55\%) attended more than 3 (out of
a total of 10) sessions (Figure~\ref{q3}). The rest (45\%) attended 3
or fewer.

\subsection{How did the online video presentations compare to conventional conference talks?}
\begin{figure}[hbt!]
\centering
\includegraphics[width=0.7\columnwidth]{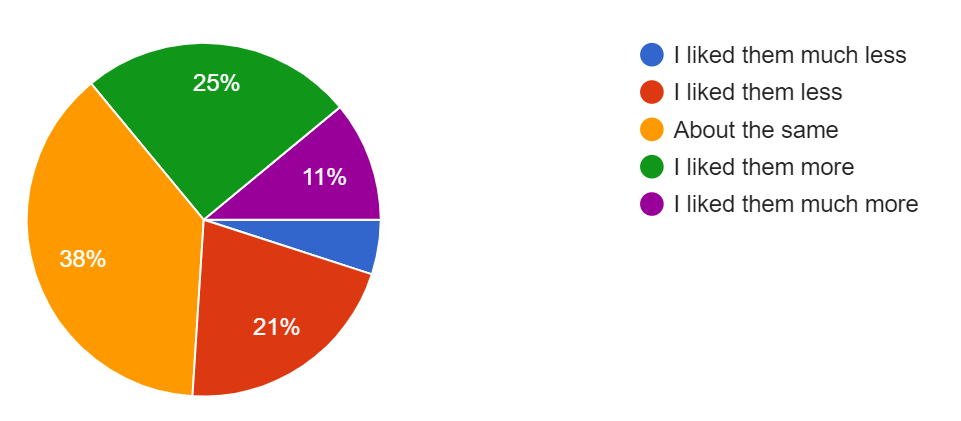}
\caption{\label{q4}How did the online video presentations compare to
                   conventional conference talks?}
\end{figure}

About one-third (36\%) of the respondents liked online video
presentations more than conventional conference presentations, about
another third (38\%) of the respondents indicated no preference, and a
minority (26\%) liked them less (Figure~\ref{q4}). We believe this is
overall a positive vote for the concept of online, pre-recorded presentations.

\subsection{What is the ideal length of a research talk for an online conference?}
\begin{figure}[hbt!]
 \centering
 \includegraphics[width=0.65\columnwidth]{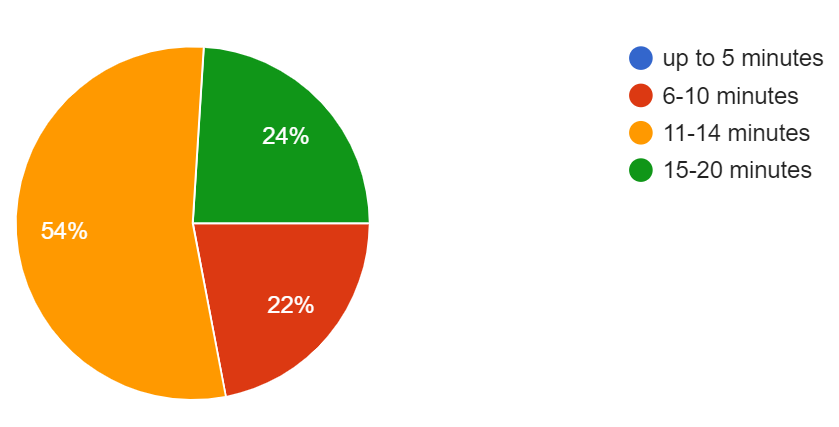}
\caption{\label{q5}What is the ideal length of a research talk for an online conference?}
\end{figure}

More than half of respondents voted for a duration of 11-14 minutes
(54\%) with an additional one-quarter (24\%) indicating a preference
for 15-20 minutes (Figure~\ref{q5}). Overall, an overwhelming majority
(78\%) indicated a preference for $>$11 minutes. There is little
support for 6-10 minutes talks (22\%), and no support at all for talks
less than 5 minutes.

\subsection{As an attendee, which kind of model(s) do you prefer?}
\begin{figure}[hbt!]
 \centering
 \includegraphics[width=0.95\columnwidth]{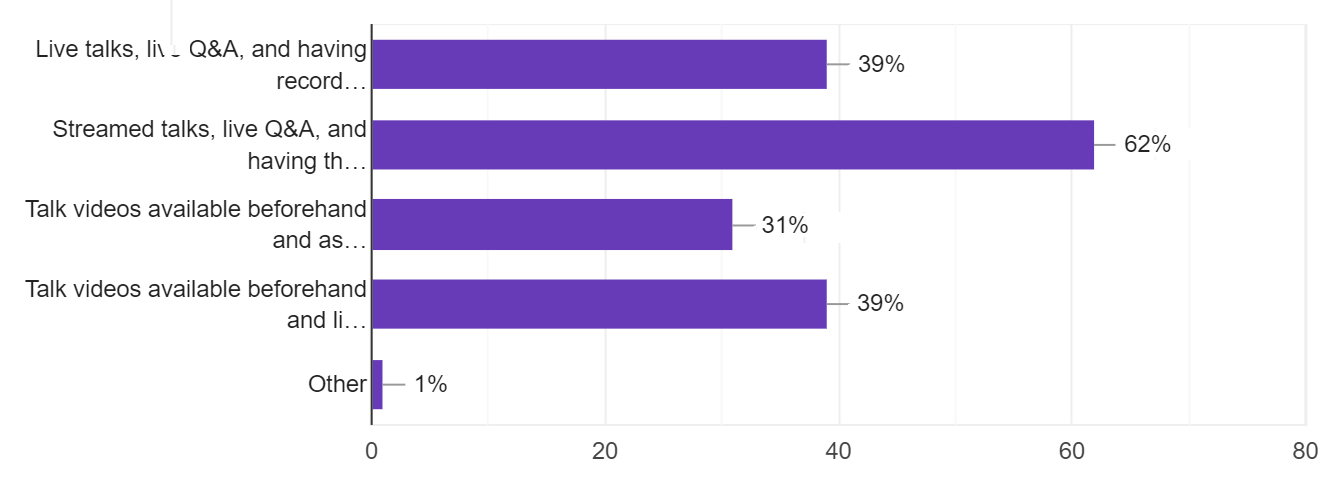}
\caption{\label{q6}Which kind of model(s) do you prefer? Multiple choices possible}
\end{figure}

Results (Figure~\ref{q6}) indicate that a majority of votes (62\%) went to the model of
\begin{itemize}
\item Streamed talks, live Q\&A, and having the videos available after the conference.
\end{itemize}

This option is similar to the one followed in EuroSys’20 since
pre-recorded talks were made available at about the time the
conference started. Next in preference with an equal amount of votes
(39\% each) were the models:
\begin{itemize}
\item Talk videos available beforehand and live Q\&A in discussion sections.
\item Live talks, live Q\&A, and having recorded talks available after the conference.
\end{itemize}

Finally, a smaller fraction of the vote (31\%) went to the option of
\begin{itemize}
\item Talk videos available beforehand and asynchronous Q\&A (e.g. over Discord or Slack).
\end{itemize}

This is an indication that attendees have a clear preference for the
synchronous streamed video and live Q\&A model. There was no stated
preference for having videos available in advance.

\subsection{Was the software infrastructure support for EuroSys’20 adequate?}
\begin{figure}[hbt!]
\centering
\includegraphics[width=0.7\columnwidth]{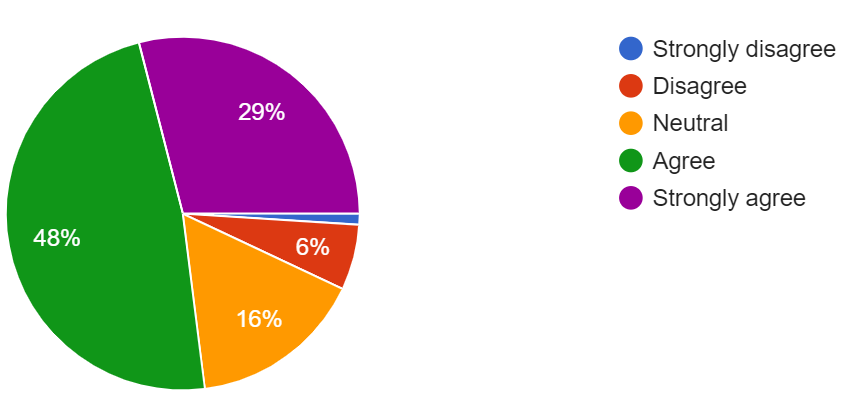}
\caption{\label{q7}Was the software infrastructure adequate for supporting EuroSys’20?}
\end{figure}

A large majority of respondents (77\%) agreed that the software
infrastructure we put in place was adequate for supporting an online
conference (Figure~\ref{q7}).

\subsection{Asynchronous interaction platforms (Slack, Discord)}
We included four questions on the use of Slack and Discord
platforms. A majority of respondents (77\%, increasing to 92\% when
counting neutral opinions) agreed that Slack was useful in
EuroSys’20. We believe that this vote indicates a preference for
asynchronous interaction platforms in general. Half of the respondents
(49\%) indicated that they had joined more than 5 Slack channels.

The majority indicated light or no use of Discord (2\% spoke in more
than 5 discussions and about 2\% listened to more than 5
discussions). The organizers noted that Discord discussion sections
during the Hallway Track were indeed used and received positive
comments by those who did. However, overall use did not match the
level of activity in the Slack workspace, which had 579 registered
(many of them active) members.

While Slack and Discord share many features and thus we believe that
the Discord platform could have been equally popular in EuroSys’20 had
it been the only asynchronous interaction platform, the fact that a
Slack workspace was available and in use several days before Discord
(while the latter was under development given very tight time
schedules), meant that it was hard to expose attendees to the full
range of capabilities of the service developed in Discord. 

From the organization point of view, there is strong belief that
Discord can significantly improve the level of
interaction between conference attendees in an online conference and
that it should be further tested in future online conferences to
evaluate its full potential.

\subsection{Did the conference need more social interaction?}
\begin{figure}[hbt!]
 \centering
 \includegraphics[width=0.7\columnwidth]{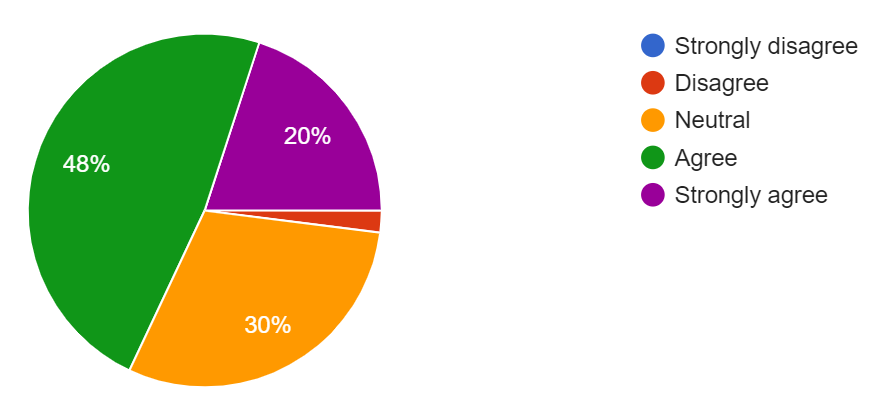}
\caption{\label{q12}Did the conference need more social interaction?}
\end{figure}

Despite the fact that attendees found the software infrastructure used
in EuroSys’20 adequate and the asynchronous interaction mechanisms
useful, a large majority (68\%, increasing to 98\% when counting
neutral opinions) agreed that more social interaction was needed
(Figure~\ref{q12}). This is an indication that more research is needed
on new ways of social interaction in online conferences (pointing to
our conclusion to further investigate new modes of interactions such
as prototyped with Discord in EuroSys’20). It is also an indication
that an online conference may in fact never be able to fully replace
the level of social interaction in a physical conference.

\subsection{Would you choose to virtually attend a physical conference?}
\begin{figure}[hbt!]
 \centering
 \includegraphics[width=0.65\columnwidth]{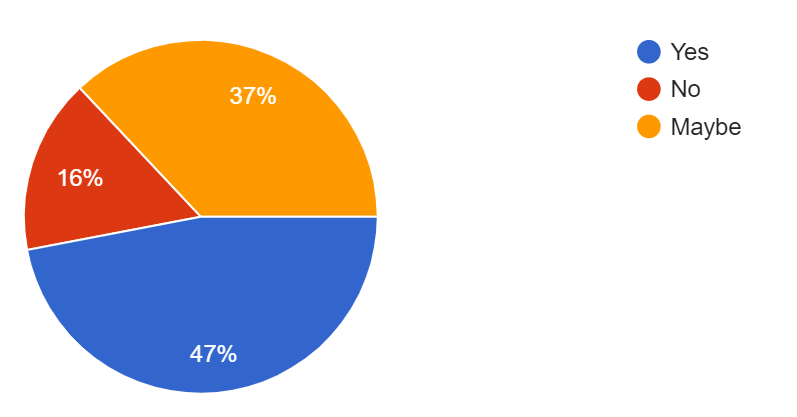}
\caption{\label{q13}Assume that EuroSys would be held physically. Would you attend virtually, if this option existed?}
\end{figure}

About half of respondents indicated that an online option such as
presented at EuroSys’20 would be a preferable option to attend a
future EuroSys, even when the option of physical attendance existed
(Figure~\ref{q13}). A large fraction of the vote (37\%) seem to be
unsure and would probably decide weighing other factors. We believe
that this result indicates that the availability of an online option
to a physical conference may have an impact on physical attendance, which should
be taken into account by conference organizers.

\subsection{Reasons to attend a hybrid conference virtually}
\begin{figure}[hbt!]
 \centering
 \includegraphics[width=0.9\columnwidth]{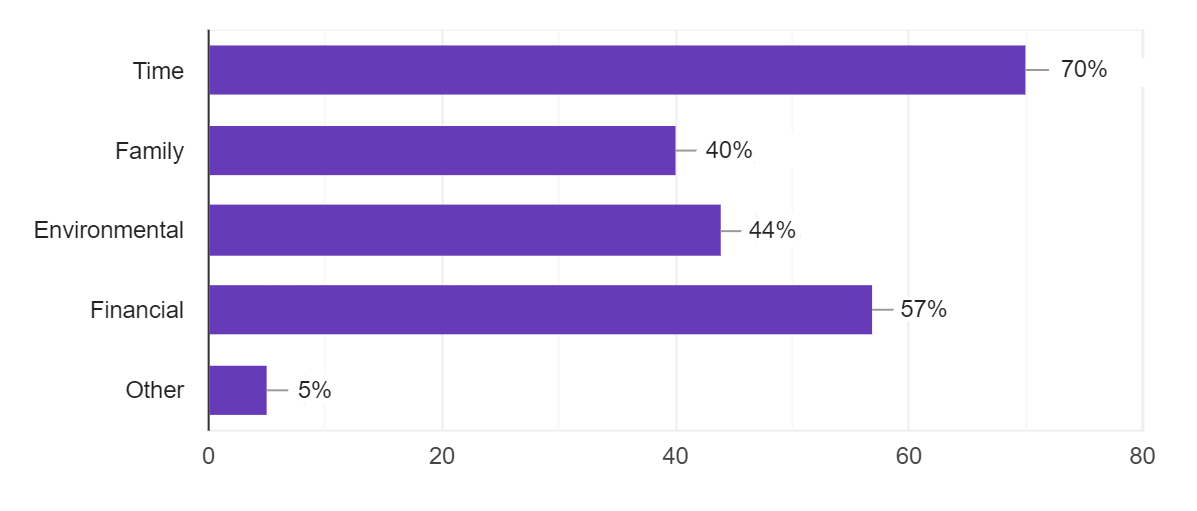}
\caption{\label{q14}For which reasons would you consider to attend a hybrid conference virtually?}
\end{figure}

Time is the most important reason for virtually attending a hybrid
(physical+online) conference for most respondents (70\%), followed by
cost (57\%), and environmental (44\%) or family (40\%) reasons
(Figure~\ref{q14}).

\subsection{Would you attend EuroSys 2021 if held virtually only?}
\begin{figure}[hbt!]
 \centering
 \includegraphics[width=0.7\columnwidth]{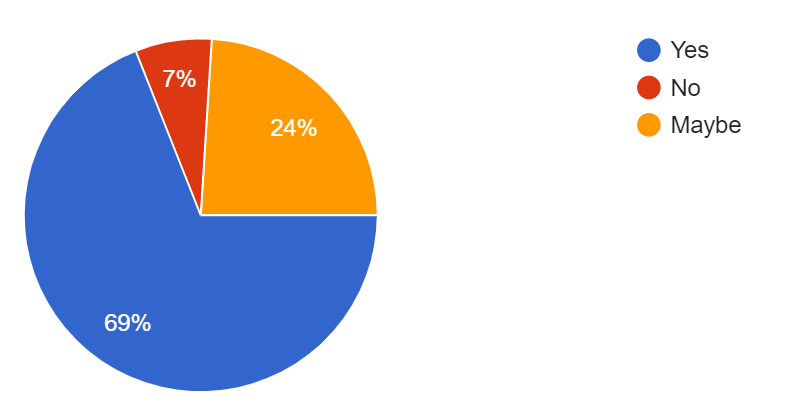}
\caption{\label{q15}Assume that EuroSys 2021 would be held virtually only. Would you attend?}
\end{figure}

A majority of the vote (about 70\%) indicate that they would attend
EuroSys’21 if held virtually only (Figure~\ref{q15}). Only a small
minority (7\%) indicate a strong negative opinion for the virtual-only
option.

\subsection{Augmenting future physical EuroSys conferences with online features}
\begin{figure}[hbt!]
 \centering
 \includegraphics[width=0.7\columnwidth]{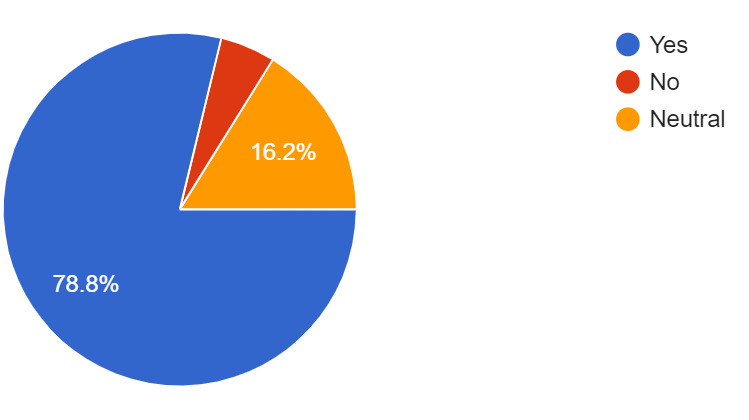}
\caption{\label{q16}Would you like to augment the future physical EuroSys conferences with online features?}
\end{figure}

The sixteenth question in our survey asked whether attendees would
like to augment future physical EuroSys conferences with some online
features used in the virtual conference (e.g., Slack or Discord
channels, short teaser videos).  A large majority of the vote in this
question indicate that attendees find asynchronous communication
channels and other online features useful, even for physical
conferences (Figure~\ref{q16}).

\subsection{Was virtual EuroSys 2020 better or worse than what you expected a virtual conference to be like?}
\begin{figure}[hbt!]
 \centering
 \includegraphics[width=0.7\columnwidth]{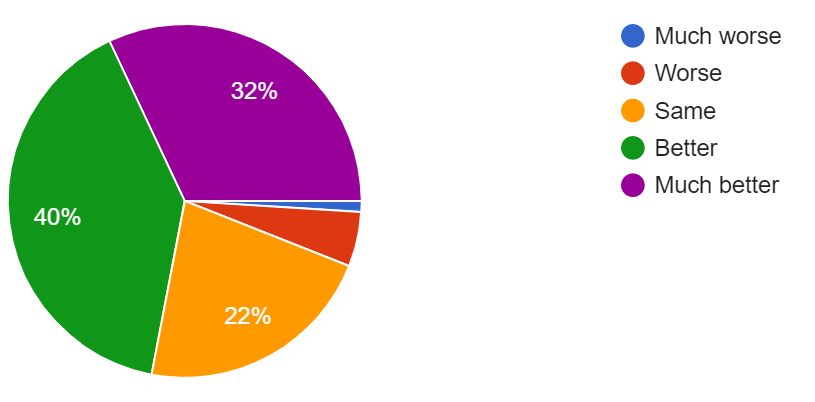}
\caption{\label{q17}Was virtual EuroSys 2020 better or worse than what you expected a virtual conference to be like?}
\end{figure}

A large majority of respondents (72\%) found virtual EuroSys’20 better
or much better than what they expected a virtual conference to be like
(Figure~\ref{q17}).

\subsection{Overall satisfaction with the organization of virtual EuroSys'20}
\begin{figure}[hbt!]
 \centering
 \includegraphics[width=0.7\columnwidth]{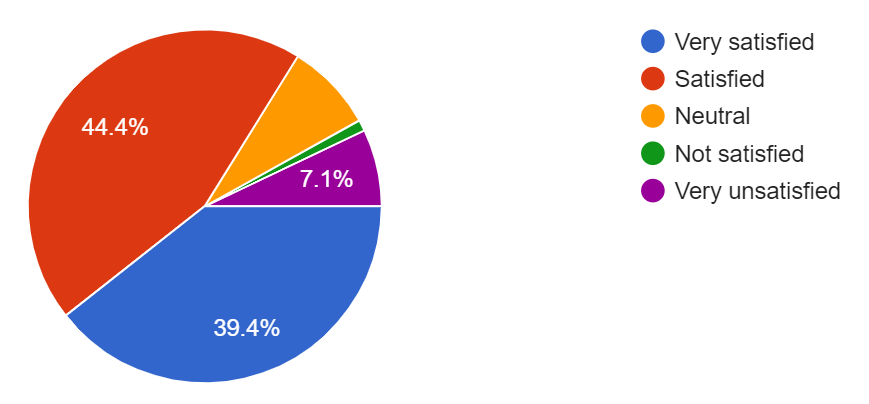}
\caption{\label{q18}How satisfied are you overall with the organization of virtual EuroSys'20?}
\end{figure}

Similar to the previous question, a large majority of respondents
(about 85\%) were satisfied or very satisfied by the organization of
virtual EuroSys’20 (Figure~\ref{q18}).

%% file: challenges.tex
\section{Lessons learned}
\label{ch:Challenges}

The following are some of the lessons we learned, and noted by
participants:

\paragraph{Timezone:} 
Although the organizers made a significant effort to spread sessions
in different timezones, US attendees could not make morning sessions
in Europe, missing nearly half the conference. Attendees from Asia on
the other hand could easily attend morning sessions in Europe.

\paragraph{Program sessions:}
In terms of program density, having to do 43 talks in 2.25 days is
challenging, especially when attending remotely.

There was positive feedback for the session format with streamed talks
and live Q\&A. 11-14 min presentations seems to be the preferred
duration. Live Q\&A was not used as extensively by attendees as we
would have hoped. However, it still gave the tone of a real-time event
and was appreciated. Although this is a more general issue beyond
virtual conferences, improving such synchronous interaction with live
Q\&A remains a challenge. The feedback suggests that asynchronous
interaction platforms were well received and Slack was used
significantly by attendees during the conference.

\paragraph{Distractions in remote participation:}
It is hard for attendees to focus on the conference program and
activities around it when still immersed in everyday life (home,
office). Two asynchronous interaction platforms (Slack and Discord)
introduced complexity for participants, the large number of channels
was at times distracting, and probably unnecessary. While it allowed
EuroSys’20 to be a live experimentation platform, which was well
received by many, we would have preferred to use a single platform. As
Discord could implement all asynchronous Slack functionality we
wanted, and provide more in the direction of synchronous functionality
(hosted virtual meetings, Hallway track), in retrospect we would have
chosen Discord as the single platform. A dense
program along with multiple interaction and information channels
created at times a sense of information overload, also for organizers.

\paragraph{Workshops:}
Workshops increase the level of difficulty as they require replicating
the setup of the main conference track, for each workshop. A question
before the conference was how to support workshops: use multiple Zoom
hosts (one for each), or a feature of Zoom large-meeting licenses
called break-out rooms (all controlled by a single host). Given that
it is easier to just replicate the model developed for the main track
for workshops, multiple Zoom hosts (one for each workshop) seemed to
be the easier way to go.

\paragraph{Registration system and process:}
Development of a registration system can take significant effort, as
well as connecting it with the rest of the conference management
processes (how to map them to Zoom invitations for different
workshops, sessions, days, etc.; how to communicate all this via email
(setup mailing lists, etc.) for multiple communication needs). One has
to consider privacy statements, code of conduct agreements,
etc. Handling “on-site” registrations was not straight-forward,
because it required human intervention to link information across
tools, so this ability was not provided in EuroSys’20. 

The registration system, as in a physical conference, can and should
interact with other aspects of the conference. We did not have the
luxury to research different registration systems and pick the one
that would work best with a virtual conference. In retrospect, being able to publish an
attendee list (with people's consent for that), which is probably more important for a virtual conference, 
would have probably improved experience, addressing the feedback that having no physical view on
the audience made it unclear who else participates.

\paragraph{Zoom master is a key role:}
The Zoom master is a very important role that may become a bottleneck:
Our Zoom master was involved in training other Zoom roles, creating
play sequences, developing the registration system, managing
registrations, and communicating with participants via mailing lists,
before and during the conference.  Although he could assign sub-tasks
to a team of volunteers, in practice his tasks were not always
parallelizable. In retrospect, it would be better to split the Zoom
master role to more than one person.

\paragraph{Training:}
Significant effort and coordination was needed for:
\begin{itemize}
\item Training of session chairs and presenters for participating in
  Zoom sessions.
\item Providing content specifications ahead of time for
  authors/speakers, in case of pre-recorded presentations.
\end{itemize}

Training will indeed be needed for any new technology that is
introduced. Because Zoom and Slack have been used in the workplace a
lot, we found that people adapted pretty well to them. Attendees were
less familiar with Discord and found it more challenging, which is
interesting, as its model is similar to Slack, with only minor
(although subtle) differences.

\paragraph{Attendance:}
Attendance was lower than anticipated based on registrations (offered
for free based on sponsor support). While registrations stood at 1100,
actual attendance peaked at 240 during opening and ranged between
100-150 after that. This was no surprise: for free, it is easy to sign
up and have no qualms about skipping. Nonetheless, it was the right
thing to do in this situation. Had we not been able to rely on
sponsorships, we could have planned to charge to cover costs. Figuring
out an appropriate registration fee remains a challenge.